%% file: main.tex
\documentclass[letterpaper,twocolumn,10pt]{article}
\usepackage{usenix-2020-09}

\usepackage[normalem]{ulem}
\usepackage{authblk}
\usepackage[frozencache,cachedir=minted-cache]{minted}
\usepackage{tikz}
\usepackage{enumitem}
\usepackage{array}
\usepackage{booktabs}
\usepackage{cleveref}
\usepackage{comment}
\usepackage{xspace}
\usepackage{listings}

\begin{document}

\date{}

\newcommand{\sys}{sqlelf\xspace}

\newcommand{\numcases}{4\xspace}

\newcommand{\eg}{e.g.,\xspace}
\newcommand{\ie}{i.e.,\xspace}
\newcommand{\etc}{etc.,\xspace}

\crefname{algocf}{algorithm}{algorithms}
\Crefname{algocf}{Algorithm}{Algorithms}
\crefformat{section}{\S#2#1#3}
\Crefformat{section}{Section~#2#1#3}
\crefformat{subsection}{\S#2#1#3}
\Crefformat{subsection}{Section~#2#1#3}
\crefformat{subsubsection}{\S#2#1#3}
\Crefformat{subsubsection}{Section~#2#1#3}
\crefformat{figure}{figure~#2#1#3}
\Crefformat{figure}{Figure~#2#1#3}

\newenvironment{smenumerate}%
  {\begin{enumerate}[itemsep=-0pt, parsep=0pt, topsep=0pt, leftmargin=2pc]}
  {\end{enumerate}}

\newif\ifcomments
\commentstrue

\definecolor{AQColor}{HTML}{000099}
\newcommand{\arq}[1]{\ifcomments\textcolor{AQColor}{[[AQ: #1]]}}
\definecolor{FZColor}{HTML}{000088}
\newcommand{\fz}[1]{\ifcomments\textcolor{FZColor}{[[FZ: #1]]}}
\definecolor{MWColor}{HTML}{000077}
\newcommand{\mw}[1]{\ifcomments\textcolor{MWColor}{[[MW: #1]]}}

\definecolor{TodoColor}{HTML}{FF5733}
\newcommand{\todo}[1]{\ifcomments\textcolor{TodoColor}{[[Todo: #1]]}}


\title{\sys: a SQL-centric Approach to ELF Analysis}

\author[1]{Farid Zakaria}
\author[1]{Zheyuan Chen}
\author[1]{Andrew Quinn}
\author[2]{Thomas R. W. Scogland}
\affil[1]{University of California Santa Cruz, Santa Cruz, CA, USA}
\affil[2]{Lawrence Livermore National Laboratory, Livermore, CA, USA}


\maketitle

\thispagestyle{empty}

\input{text/abstract}
\input{text/introduction}

\input{text/object_formats}

\input{text/data_model}
\input{text/sqlelf}
\input{text/evaluation}
\input{text/related}
\input{text/conclusion}

\bibliographystyle{plain}
\bibliography{references}

\end{document}

%% file: text/abstract.tex
\begin{abstract}
The exploration and understanding of Executable and Linkable Format~(ELF) objects underpin various critical activities in computer systems, from debugging to reverse engineering. Traditional UNIX tooling like readelf, nm, and objdump have served the community reliably over the years. However, as the complexity and scale of software projects has grown, there arises a need for more intuitive, flexible, and powerful methods to investigate ELF objects. In this paper, we introduce \sys, an innovative tool that empowers users to probe ELF objects through the expressive power of SQL. By modeling ELF objects as relational databases, \sys offers the following advantages over conventional methods:
\begin{description}[noitemsep]
    \item[Expressive Queries] The inherent structure and expressiveness of SQL allow for intricate, multi-dimensional queries, enabling users to obtain insights previously impractical or cumbersome with traditional tools.
    \item[Aggregation] Easily aggregate data, offering a holistic view of properties and relations across many ELF objects.
    \item[Joining and Relating Data] Seamlessly relate different sections or parts of a single or multiple ELF files, providing a unified view of the interplay between them.
    \item[Extensibility] The relational database model is amenable to extensions, allowing for easy integration with other tools and datasets.
    \item[Accessibility] SQL, being a widely known language, lowers the entry barrier for newcomers, making ELF exploration accessible to a broader audience.
\end{description}
Our evaluations demonstrate that \sys not only provides more nuanced and comprehensive insights into ELF objects but also significantly reduces the effort and time traditionally required for ELF exploration tasks.
\end{abstract}

%% file: text/introduction.tex
\section{Introduction}

Software projects pervasively depend on external components, also called software dependencies.  For example, in 2019, Sonatype found that the average application includes hundreds of software dependencies~\cite{sonatype19}.  Developers turn to software dependencies because the approach improves their productivity~\cite{Haefliger08, Tracz88};  a developer can use an existing well-tested library as a module in their project instead of writing their own implementation. 

While the growth of software dependencies simplifies software development, it greatly complicates system administration. Systems maintain software dependencies for compiled software (\eg C, C++, Rust, \etc) as shared libraries stored in conventional file-system locations.  To run an application, a dynamic loader transitively traverses the application's dependencies, and the application's dependencies' dependencies (and so on), following a myriad of specifications and conventions to decide which shared library to load.  Given the number and complexity of the relationships between software dependencies, it is difficult for system administrators to maintain system health.  Consequently, today's sysadmins struggle to maintain a consistent rationalized view of their systems and more precisely understand the binary files that comprise them.

For example, many Linux distributions enforce single-version policies such as the \emph{One-Definition Rule}~\cite{Definiti47:online} for most software that limit the opportunities for unsolvable conflicts between dependencies, popularly known as \emph{dependency-hell}, that emerge when executables pull in multiple copies of the same dependency at different versions. Nevertheless, there are often exceptions to this rule; for example, libraries pursuing a large version migration, such as Python and GTK, permit concurrent versions. These exceptions have caused bugs in some of the most widely used Linux distributions~\cite{GentooGdbBug:online}. Even aside from such exceptions, the single-version policy lacks consistent, practical, large-scale enforcement that no two shared libraries export symbols of the same name.

The fundamental issue is that system administrators lack the tools necessary to analyze of their systems.  Using current state-of-the-art object code inspection tools (\eg \verb`readelf`, \verb`nm`, and \verb`objdump`), a user can \emph{observe} the individual object metadata and code for each software package, but not \emph{analyze} them.  Moreover, existing tools do not allow an administrator to analyze the metadata of multiple objects.  Consequently, system administrators rely on their own manually written, ad-hoc scripts to investigate their systems, such as to track dependencies and diagnose unexpected behaviors. These scripts, while critical, are easy to get wrong.

This paper argues for a new approach to maintaining software dependencies that views software dependency administration from the lens of data analytics.  We envision that each system will maintain a datastore that exposes all software resident on the system.  With the datastore, system administrators will have visibility into the code, metadata and dependencies of their entire system, which they can succinctly analyze to develop and maintain a consistent view of overall system health.

To realize our vision, we first survey the object file formats (\eg Executable and Linkable Format (ELF), Mach-O, and PE) and object code inspection tools used in existing systems (\cref{sec:object-formats}).  Our survey provides three insights: First, the (meta)data contained in object file formats is, in essence, relational data.  Second, while there are small differences across the object formats, they are similar enough to be visualized through the same underlying data model.  Third, existing code inspection tools are inadequate for many system administrator tasks.

Based upon our insights, we design a practical data model (\cref{sec:data-model}).  The data model exposes all metadata information and code of all software in a system as a relational database.  The data model largely mirrors the structure of ELF, diverging only in instances where the existing ELF structure hampers analysis. As a relational database, the data model supports analysis via SQL queries. For example, a system administrator can write a SQL query that determines whether a library update would shadow symbols that are already exported by libraries in their system, a common cause of dependency-related software bugs through violation of the \emph{One Definition Rule}.

We realize the relational software dependency model with \sys (\cref{sec:design}).  \sys is a prototype system built for Linux that exports information from ELF files into a SQLite relational database.  Thus, \sys benefits from SQLite's built-in support for caching, indexing, and transactions and external tooling for visualization from enterprise grade solutions such as Tableau~\cite{Tableau:online} and Power BI~\cite{DataVisu17:online} or emerging open-source options such as Datasette~\cite{Datasett26:online} which simplifies publishing SQLite databases for exploration.

We use the relational model and \sys for \numcases software case studies:
\begin{smenumerate}
    \item \emph{auditwheel}: We reimplement the Python binary dependency policy tool auditwheel in terms of \sys.
    \item \emph{dynamic linking}: We provide an implementation of the musl dynamic-loader leveraging the \sys data model.
    \item \emph{symbol interposition}: We demonstrate how SQL simplifies the common sysadmin task of determining whether two symbols might shadow each other.
    \item \emph{aggregate analysis}: We reimplement \verb`elf_diff`, a tool to compare multiple ELF binaries, in terms of \sys. Additionally, we create single-file databases that comprise a complete Linux distribution and demonstrate how \sys simplifies analysis.
\end{smenumerate}
In each case we demonstrate a reduction in code and improvement in comprehensibility.

In sum, our contributions are as follows
\begin{smenumerate}
\item A relational data model for exposing the relationships, including but not limited to the dependencies between all software libraries and applications in a deployed system.
\item \sys, a realization of this whole-system relational model for Linux and ELF binaries. 
\item An evaluation of the relational data model and \sys on \numcases case studies, exemplifying the potential for the tool to common but cumbersome software analysis.
\end{smenumerate}

The rest of this paper is as follows:  We survey existing object formats and software dependency management tools (\cref{sec:object-formats}).  Then, we describe the relational data model (\cref{sec:data-model}).  We discuss the design of \sys (\cref{sec:design}), a realization of the data model specific to Linux and ELF object files. Next we evaluate \sys's efficacy by using it in \numcases case studies (\cref{sec:eval:case-studies}) and evaluate \sys's performance  on those case studies (\cref{sec:evaluation}).  We discuss related work (\cref{sec:related}) and conclude (\cref{sec:conclusion}).

%% file: text/object_formats.tex
\section{Object File Formats}\label{sec:object-formats}
An object file is a file that holds the machine code output generated by an assembler or compiler. Multiple formats of object files exist, each tailored for specific purposes and operating systems. These files play a role in both program linking and program execution. Among the most well-known object formats are: Mach-O which is utilized in macOS, the Executable and Linkable Format (ELF) which is predominantly used in Linux and Unix-like operating systems, and Portable Executable (PE) which is the file format used by Microsoft Windows. 

In any variant, executable programs encompass the following components:
\begin{itemize}[noitemsep,topsep=0pt]
    \item \textbf{Program text region}: a contiguous listing of instructions.
    \item \textbf{Program data region}: a space for data with predefined values.
    \item \textbf{Associated information region}: details about external variables.
    \item \textbf{Header region}: housing the positions of the above information and its related metadata.
\end{itemize}

\subsection{ELF}
The ELF (Executable and Linkable Format) object file format has dominated the Linux space as the object format in which all executables, relocatable files and shared object files
are created since 1999 when it was chosen as the standard binary file format for Unix and Unix-like systems. In fact, as of 2022 (Linux kernel version 5.18), support for the predecessor object file format \verb`a.out` was removed leaving ELF as the defacto object file format.

\begin{figure}[t]
\includegraphics[width=\columnwidth]{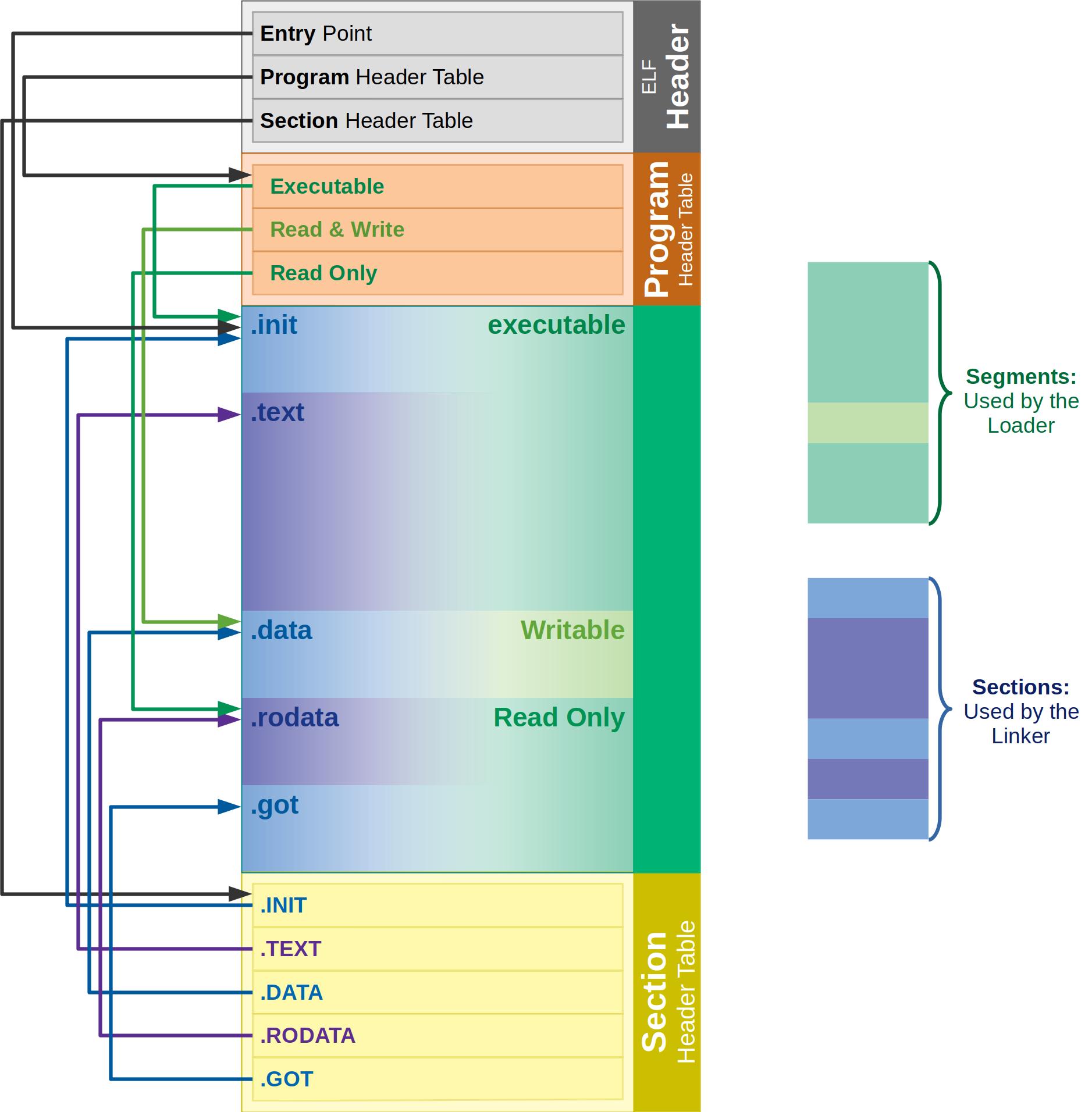}
\caption{A simplified view of the ELF layout and its duality for execution and linking. \copyright 
 A. Burtsev}
\label{fig:elf-containers}
\end{figure}

The ELF object format helped solve many of the problems faced with prior object formats that increased as a result of our heterogeneous world such as varying ISA, ABI, and byte orderings.
ELF was designed to be adaptable to different architectures and to support enhancements without breaking compatibility. It is a highly configurable specification that has over time become
more dependent on shared conventions, however the simplest comprehension of the file format is a series of \emph{containers} and a header table that provides access to them by name as depicted by Figure~\ref{fig:elf-containers}. The format is comprised of either \emph{Sections} if the file is used as a relocatable object; otherwise the file format contains \emph{Segments} if the file is an executable or shared object file to be used by the loader (operating system or dynamic linker). Complicating matters further is that oftentimes an executable or shared object file will contain both sections and segments with a mapping between the two.

The linking view, represented through sections in an ELF file, primarily caters to the compilation and linking stages of binary creation. Each section holds information pertinent to these stages, such as symbol tables, relocation entries, and debug information. The granular categorization of data into \emph{named} sections, by convention, such as \emph{.text} for program instructions, \emph{.data} for initialized data, and \emph{.bss} for uninitialized data, among others, facilitates linkers and compilers in efficiently managing, modifying, and linking code and data during the build process.

Conversely, the execution view, represented through segments, is crafted with a lens focused on the loader and the runtime system. A segment, in the context of ELF, is a contiguous block of memory that encompasses one or more sections, designed to be loaded into memory by the system loader via \verb`mmap`. The execution view abstracts away the granular details present in the linking view, instead providing a coarser, more streamlined perspective that is optimized for efficient loading and execution.

The bifurcation of views in ELF files balances the often conflicting requirements of compile-time flexibility and run-time performance. The linking view, with its detailed and structured sections, provides a rich environment for linkers and compilers to perform optimizations, manage symbols, and facilitate debugging. On the other hand, the execution view, with its coarser segments, ensures that the runtime system can load and execute binaries with minimal overhead, ensuring efficient utilization of system resources.

The duality, although pragmatic in certain contexts, introduces complexities and potential inefficiencies that merit exploration. A minor, but notable improvement was introduced in the Solaris 11.4 (2008) operating system. Traditionally, only the sections in the linking view have a text name to help tools and users to distinguish their use by convention such as \emph{.data}, \emph{.text} and \emph{.rodata}. Program headers were left nameless perhaps due to the fact that historically there typically existed only one to two segments but today there are many mappings per object. For the additional 32-bit word per program header and storing the associated strings in the string table, Program Headers in Solaris now have associated names which take much of the guesswork out of discovering what a given segment is for~\cite{ELFProgr58:online}.

Over the years, this landscape has seen a consistent reliance on a set of traditional UNIX tools like \verb`readelf`, \verb`nm`, and \verb`objdump` that simply dump raw ASCII text to the console such as is shown in Figure~\ref{fig:readelf-sample}. While these tools have stood the test of time, and one could argue they adhere to the Unix philosophy of doing one thing well~\cite{mcilroy1978unix}, this same philosophy leads to opaque combination of commands to extract meaningful answers to questions one may have about the data. This status quo for investigating binaries has largely gone unchallenged. In an era where data is growing at an unprecedented scale, and software projects have exploded in size and complexity, there's a growing need to re-evaluate our toolset.

\begin{figure}[ht]
\begin{minted}[frame=single,fontsize=\footnotesize]{console}
$ readelf --demangle --dyn-syms /usr/bin/ruby

Symbol table '.dynsym' contains 22 entries:
Num:  Value Size Type    Bind   Vis   Ndx    Name
 0:    0     0 NOTYPE  LOCAL  DEFAULT  UND 
 1:    0     0 FUNC    GLOBAL DEFAULT  UND ruby_run_node
 2:    0     0 NOTYPE  WEAK   DEFAULT  UND __gmon_start__
 3:    0     0 FUNC    GLOBAL DEFAULT  UND ruby_init
 4:    0     0 FUNC    GLOBAL DEFAULT  UND ruby_options
 5:    0     0 NOTYPE  WEAK   DEFAULT  UND _ITM_deregis
 6:    0     0 NOTYPE  WEAK   DEFAULT  UND _ITM_registe
\end{minted}
\caption{Sample output of present-day tooling to introspect dynamic symbols within an ELF file.}
\label{fig:readelf-sample}
\end{figure}

\subsection{Other file format}
\begin{figure}[ht]
\centering
\includegraphics[scale=0.5]{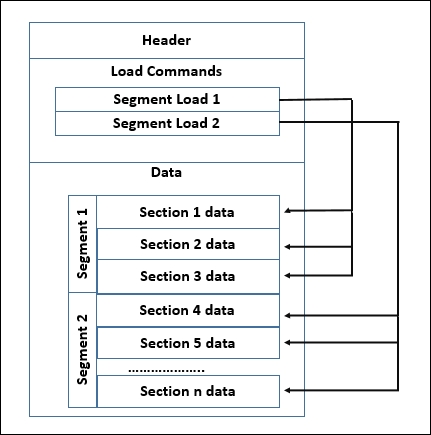}
\caption{A simplified view of the Mach-O layout, adapted from "Mobile Application Penetration Testing"
by Vijay Kumar Velu (2016), as cited in O'Reilly (2023)}
\label{fig:macho}
\end{figure}

\begin{figure}[ht]
\centering
\includegraphics[scale=0.8]{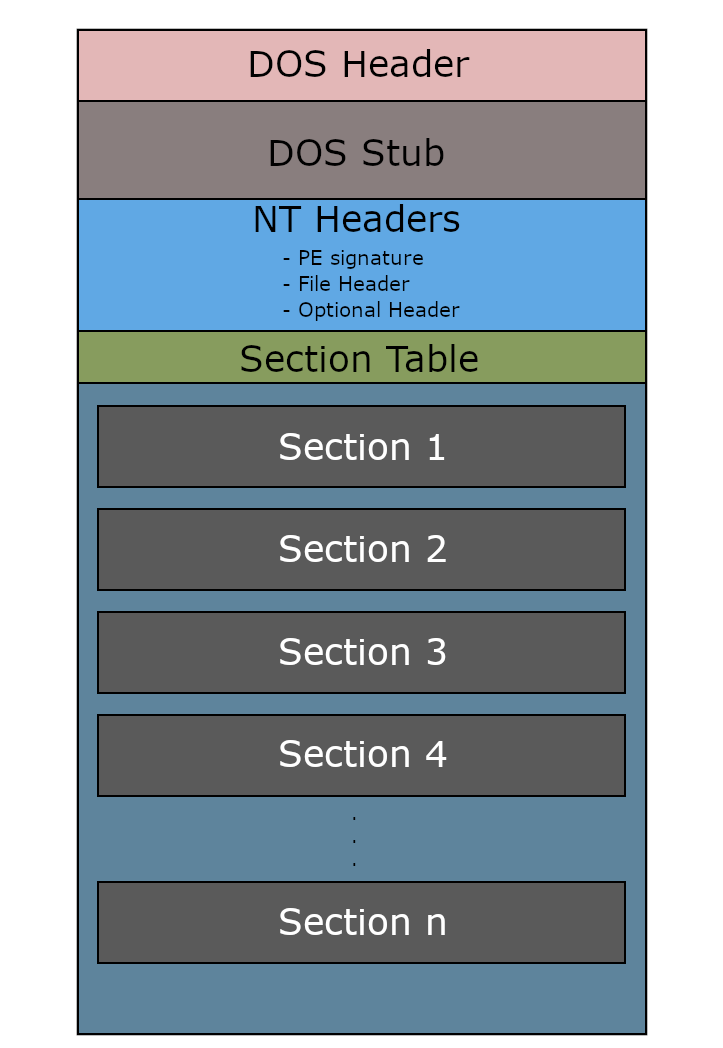}
\caption{A simplified view of the PE layout, adapted from "A dive into the PE file format - PE file structure"
by Ahmed Hesham (Oct 22, 2021)}
\label{fig:pe}
\end{figure}

Mach-O and PE are native object file format in macOS and Windows OS respectively, and they both serve the similar purpose as ELF and share many fundamental features. The simplified layout of Mach-O and PE are illustrated by Figure~\ref{fig:macho} and Figure~\ref{fig:pe} separately. 

Mach-O, and PE begin with a header that defines machine-specific attributes and provides essential information about the binary’s architecture and characteristics. This header enables the loader to understand, interpret, and execute the file.

Each format has its own nuances for storing program (meta)data:
\begin{itemize}
    \item Mach-O files, used primarily in macOS, have a header followed by load commands that define the file structure and linking attributes. These commands specify the file’s layout in memory, symbol table locations, execution state, and linked libraries. Mach-O files also use segments, like \emph{\_\_TEXT} and \emph{\_\_DATA}, each containing multiple sections for organizing data, and include a field for setting memory protection levels.
    
    \item PE files, common in Windows, start with a DOS header for backward compatibility, followed by a DOS stub and an NT header. The NT header comprises a PE signature, \emph{IMAGE\_FILE\_HEADER} for metadata (like architecture and file attributes), and an \emph{IMAGE\_OPTIONAL\_HEADER} containing essential information for the PE loader. Sections in PE files, defined by Section Headers, contain the actual data and code of the executable.
\end{itemize}

Despite their differences, ELF, Mach-O, and PE formats share the fundamental concept of using sections to store different types of data and code. These sections allow for a structured representation of a binary's components.

In terms of memory loading and data organization:
\begin{itemize}[noitemsep,topsep=0pt]
    \item ELF relies on \verb|Program Headers| to define the attributes and layout of segments.
    \item PE uses the \verb|IMAGE_OPTIONAL_HEADER| and section data for similar purposes.
    \item Mach-O employs \verb|Segment Load Commands|, serving a function akin to ELF’s \verb|Program Headers|.
\end{itemize}

The similarities between these three formats reflect common requirements in binary formats across different operating systems.

%% file: text/data_model.tex
\section{Data Model}\label{sec:data-model}

The usefulness of a declarative language is bounded by the data model upon which it's exercised. A key insight is that there exists a data-model that is \emph{isomorphic} to the current-day binary formats. The simplest way in which to demonstrate this equivalence is to imitate this domain model when building the corresponding relational data model.

\begin{figure}[t]
    \centering
    \includegraphics[scale=0.3]{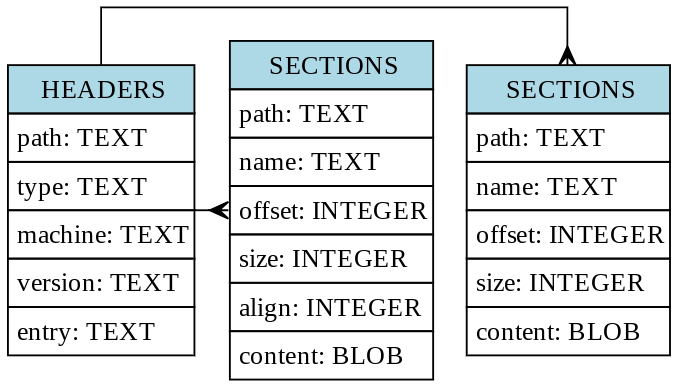}
    \caption{An equivalent but non-useful data model to the ELF file format.}
    \label{fig:minimal-schema}
\end{figure}

We chose to model the ELF file format as a relational model due to our personal prior experience with the format. At the most congruent mapping, there would exist 3 tables: the ELF header, segments and sections tables with much of the \emph{interesting bits} of the object model packed away in an opaque \verb`BLOB` column as shown in Figure~\ref{fig:minimal-schema}.

\begin{figure*}[ht]
    \centering
    \includegraphics[scale=0.28]{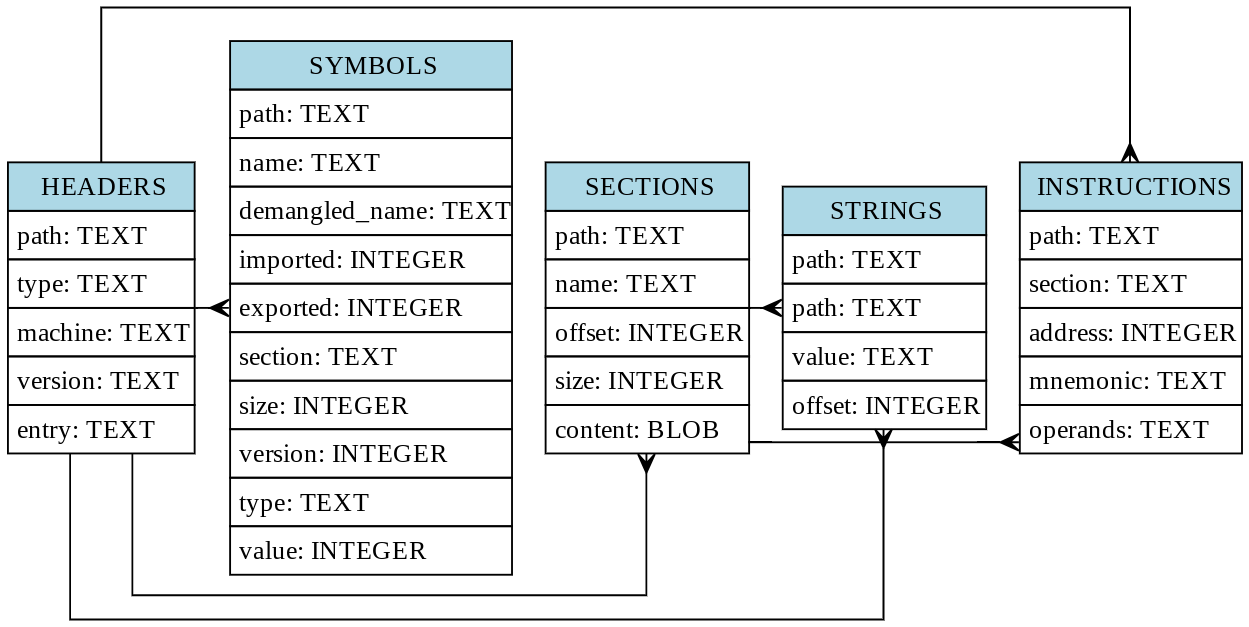}
    \caption{A portion of the schema available to be queried.}
    \label{fig:sqlelf-schema}
\end{figure*}

In order to provide a \emph{useful} relational data-model that can support a wide array of analyses, further enhancements to the model can be made by elevating the contents of certain well known sections into their own unique table, such as instructions, strings and symbols. Segments need no further lifting as they remain identically unstructured within the ELF file and serve only to be a region of contiguous memory which can easily be \verb`mmap` into process address space. A portion of this augmented data model can be seen in Figure~\ref{fig:sqlelf-schema}.

We chose to also elevate many common expressions into singular columns to simplify analysis against the table. For example, for a symbol to be considered \verb`exported` it must meet multiple criteria: must reference a section other than \verb`SHN_UNDEF` (undefined section), have an address, and be of a specific symbol type and binding. Rather than having each query articulate these constraints when searching for exported or imported symbols, which is a common inquiry for ELF objects, the user of the data model can now leverage and filter against a simple column.

The opportunities to extend and simplify the data model are endless and often more in service of the analysis one is performing. Our data model includes several such simplifications that we found desirable when applying \sys to the \numcases case studies outlined in Section~\ref{sec:eval:case-studies}. Any missing abstractions to the model however can be easily added using the notion of a SQL view, which allows one to define virtual or persistent data model abstractions from the result of a query (e.g., \texttt{CREATE VIEW view\_name AS SELECT column1, column2, ... FROM table\_name WHERE condition})

Relational databases are well known for their ability to explicitly represent and navigate relationships, bringing clarity and structure to data that might otherwise seem opaque and unconnected. One of the more challenging aspects of modeling the ELF object in a relational schema is the representation of the dynamic (\verb|.dynamic|) section. The dynamic section in ELF files is particularly intricate due to its nature of requiring context-specific interpretation. Each entry in this section holds a value (\verb|d_tag|) that dictates its relationship, significance, and how to interpret the remaining data. The value of the remaining data may be hierarchical in nature and point to an offset within other sections. While SQL has enough primitives to handle this dynamicism, it requires the user to know a substantial amount of the internals of ELF. We found that the correct abstractions can make working with the dynamic entries much more pleasant, keeping the leakiness of the underlying object model minimal and without having to create a unique table for each unique entry type. Figure~\ref{fig:sql-dynamic-entris} demonstrates how a SQL query to determine the \verb|RUNPATH| for a particular file can be radically simplified simply by offering a table, \verb|elf_strings|, that contains relational-model equivalence for every string found in the various string tables (e.g., \verb|.symtab|, \verb|.strtab| \& \verb|.dynstr|).

\begin{figure*}[ht]
\centering
\begin{minipage}[b]{.49\textwidth}
   \centering
    \begin{minted}[frame=single,fontsize=\footnotesize]{sql}
SELECT 
    SUBSTR(
        untrimmed, 
        1, 
        INSTR(untrimmed, char(0)) - 1
    ) 
FROM (
    SELECT CAST(
        SUBSTR(elf_sections.content,
               elf_dynamic_entries.value +1 ) as char
        ) AS untrimmed
    FROM elf_sections, elf_dynamic_entries
    WHERE elf_dynamic_entries.tag = 'RUNPATH' AND
    elf_sections.name = '.dynstr'
    )
    \end{minted}
\end{minipage}\hfill
\begin{minipage}[b]{.49\textwidth}
   \centering
    \begin{minted}[frame=single,fontsize=\footnotesize]{sql}
SELECT elf_strings.value
FROM elf_dynamic_entries
INNER JOIN elf_strings
    ON elf_dynamic_entries.value = elf_strings.offset
WHERE elf_dynamic_entries.tag = 'RUNPATH'
\end{minted}
\end{minipage}
\caption{Demonstrating how SQL queries against the data model can be radically simplified through the right abstractions.}
\label{fig:sql-dynamic-entris}
\end{figure*}

%% file: text/sqlelf.tex
\section{\sys: Design \& Implementation}\label{sec:design}

Declarative languages have long held a significant place in data management systems, epitomized by the widespread adoption of SQL by nearly every relational database system in existence. The essence of a declarative language like SQL lies in its ability to express the logic of computation without describing its control flow, focusing on the \emph{what} rather than the \emph{how}. This abstraction from implementation details enables one to specify the data they want to retrieve or manipulate without outlining the specific steps the system must take to perform the task. This high level of abstraction is particularly advantageous in data management systems, allowing both efficiency and accessibility by delegating optimized execution to the query engine.

Second to the venerable \emph{file}, the idea in the Unix \& Linux that everything is a file, the most common and important unit of data management on these systems are executables and shared object files stored in the Executable and Linkable Format (ELF). Despite their prevalence, introspection to the binary format has long been consigned to a small cohort of tools, such as \verb`readelf` and \verb`objdump`, that offer limited capability to perform advanced analysis beyond merely dumping a textual encoding of the format. The problem is more bleak for those who wish to programmatically access
information of the file format as the format can be rather tricky to navigate for the reasons listed in Section~\ref{sec:object-formats}.

The pairing of battle-tested and well understood data-management principles with object file formats is a match made in heaven. We have developed \emph{\sys}, an open-source SQL query engine for the Executable and Linkable Format. \sys exposes a relational schema that either mirrors portions of the ELF file format itself or augments the format by offering \emph{virtual tables} to make advanced analysis simpler. \sys provides the following features:
\begin{itemize}[noitemsep]
    \item Provides a relational schema that mirrors the ELF file format;
    \item Augments the format with tables that simplify advanced analysis;
    \item Offers a SQL front-end API and CLI to interface with those tables;
    \item Allows exporting the data model as SQLite database files;
    \item Available as open-source MIT licensed software.
\end{itemize}


SQLite, developed by Dr. Richard Hipp in August 2000, is said to be the most widely used database engine~\cite{gaffney2022sqlite}. Its ubiquity is underscored by its pervasive deployment in various software stacks, including mobile devices, embedded systems, and numerous applications, making it arguably the world's most widely used piece of software, let alone database engine.

Unlike traditional relational database management systems (RDBMS), SQLite diverges in several key aspects that have been instrumental in its widespread adoption:
\begin{itemize}
    \item \textbf{Serverless Architecture:} SQLite operates without a separate server process, interacting directly with storage files or in-memory databases.
    \item \textbf{Single Database File:} All the data, tables, indices, and schemas are stored in a single cross-platform file, simplifying management, backup, and transfer of databases.
    \item \textbf{In-Memory Databases:} SQLite supports in-memory databases, facilitating rapid data access and manipulation without persistent storage.
    \item \textbf{Zero Configuration:} SQLite does not require configuration, making it straightforward to integrate and use in applications.
\end{itemize}

One of the compelling features of SQLite is the virtual table mechanism~\cite{TheVirtu43:online}, which allows for the definition of interfaces to custom data sources as if they were tables in a SQLite database. Virtual tables can be used to represent data structures in applications, external data sources, computed or synthetic data. This functionality is leveraged by \sys, providing a structured and queryable interface over ELF files. The virtual table mechanism enables \sys to represent various sections and structures within ELF files as tables in an SQLite database depicted in Figure~\ref{fig:sqlelf-schema}, thereby allowing users to interact with and query ELF files using SQL, blending the structured and expressive query capabilities of SQL with the rich and complex data structures found in ELF files.

While a nearly identical analogue of the schema could be used to match the underlying file format, the abstraction easily lends itself to schema augmentations that improve introspection and are only paid for at query time. Two useful examples we've found: including the disassembly for each instruction; and a table that contains all strings from each respective string table section (\emph{.strtab}, \emph{.shstrtab} and \emph{.dynstr}). While we believe we've captured practical domain fundamentals in the data model, users can augment it with SQL views according to their needs.

Virtual tables only exist for the lifetime of the database and require some additional setup and wiring. This has the added benefit users only pay for the data they query; conversely, repetitive \verb`SCAN` of the data can be expensive for complicated queries. The data and setup however can easily be persisted into the database itself using a \emph{Create Table As Select} (CTAS) statement such as \texttt{CREATE TABLE ELF\_SYMBOLS AS SELECT * FROM ELF\_SYMBOLS\_VIRTUAL}, effectively memoizing ELF file parsing. We've designed \sys so that the user can choose whether to memoize a given table on startup into an in-memory SQLite database or to lazily evaluate it during query execution. If a user knows they won't be querying one table but querying another often, this choice allows them to balance startup time with the runtime of their repeated queries. This becomes an important feature to improve overall performance as we'll see in Section~\ref{sec:evaluation}. The persistence of the virtual tables into a standard SQLite database furthermore enables the use of the broad set of SQLite tools developed by its community.

%% file: text/evaluation.tex
\section{Evaluation}
\label{sec:evaluation}\label{sec:eval}

This section describes our evaluation of \sys.  We answer two research questions: 

\begin{smenumerate}
    \item ``How does \sys and its data model simplify object maintenance for system administrators?''
    \item ``How does \sys's performance compare to existing tools for observing software objects?''
\end{smenumerate}

\subsection{Case Studies}\label{sec:eval:case-studies}

We describe \numcases case studies that use \sys and its data model to simplify object maintenance.  Each scenario describes an existing object maintenance task and outlines how a system administrator could employ \sys and current tools for the task.

\subsubsection{auditwheel}

Python is infamous for its packaging story, or lack thereof. Initially, the \emph{egg} format emerged as a prominent mechanism for distributing Python packages. These eggs encapsulated package metadata and dependencies, providing a means for dependency resolution. However, despite their widespread use, eggs were never adopted as a formal standard in the Python community. Recognizing the need for a more standardized and stable binary distribution format, the Python community introduced the \emph{wheel} format through PEP (Python Enhancement Proposal) 427~\cite{holth2012pep}. However, distributing pre-compiled shared libraries, such as Python C-extensions, via the wheel was problematic on Linux. The wheel's metadata did not include necessary information about any shared libraries' dependencies and how to retrieve them. In a rather pragmatic solution, the community put forward PEP 513~\cite{mcgibbon2016platform} which outlined a series of policies embedded shared libraries must adopt to ensure widespread compatibility (e.g., depend on a widely-compatible kernel ABI, only depend on an extremely limited set of well known shared libraries with strict compatibility policies and so forth). Any wheel which adheres to the policy is then deemed to work on most Linux distributions and is thus called \emph{"manylinux"}.

To help authors adhere to the \emph{manylinux} specification, the \emph{auditwheel} tool was made available. The tool inspects all of the ELF files inside a wheel to check for dependencies on versioned symbols or external shared libraries, and verifies conformance with the relevant manylinux policy. It also has support to repair non-conforming wheel. The tool is written in Python and requires analyzing ELF files making it a great candidate for use with \sys.

\begin{figure*}[ht]
\centering
\begin{minipage}[b]{.49\textwidth}
   \centering
    \begin{minted}[frame=single,fontsize=\footnotesize]{python}
def elf_is_python_extension(elf_file, mod_name):
    module_init_f = {
        "init" + mod_name: 2,
        "PyInit_" + mod_name: 3,
        "_cffi_pypyinit_" + mod_name: 2,
    }
    sect = elf_file.get_section_by_name(".dynsym")
    if sect is None:
        return False, None
    for sym in sect.iter_symbols():
        if (
            sym.name in module_init_f
            and sym["st_shndx"] != "SHN_UNDEF"
            and sym["st_info"]["type"] == "STT_FUNC"
        ):
            return True, module_init_f[sym.name]
    return False, None
    \end{minted}
\end{minipage}\hfill
\begin{minipage}[b]{.49\textwidth}
   \centering
    \begin{minted}[frame=single,fontsize=\footnotesize]{python}
def elf_is_python_extension(sql_engine, mod_name):
    sql = f"""
        SELECT
            CASE name
            WHEN 'init{modname}' THEN 2
            WHEN 'PyInit_{modname}' THEN 3
            WHEN '_cffi_pypyinit_{modname}' THEN 2
            ELSE -1
            END AS python_version
        FROM elf_symbols
        WHERE name IN ('init{modname}', 
                       'PyInit_{modname}',
                       '_cffi_pypyinit_{modname}')
              AND exported = TRUE
              AND type = 'FUNC'
        LIMIT 1
            """
    results = list(sql_engine.execute(sql))
    if len(results) == 0:
        return False, None
    return True, results[0]["python_version"]
\end{minted}
\end{minipage}
\caption{An auditwheel function that determines the Python version of the C extension and its SQL analog.}
\label{fig:auditwheel-code}
\end{figure*}

Every function which accesses the ELF data-structure was replaced with a SQL analog, an example of which is depicted in Figure~\ref{fig:auditwheel-code}. These changes replaced sequences of functions composed of ad-hoc logic with single SQL queries that can be easily reused across programming languages, implementations and even used interactively via the \sys command-line interface. 

\subsubsection{dynamic linking}

Object files are the interchange data format between build tools such as compilers and linkers. At an abstract level, the file format needs to contain all relevant information necessary to ultimately emit a blob of data that can be set as the entry point to a program counter and executed. The opportunity to deploy a SQL-centric approach natively to the build toolchain is an active area of investigation we are researching. 
The tail end of the build tool chain however is an area we more immediately explored, the dynamic loader.

A dynamic loader, also known as a dynamic linker, is a system utility that loads and links the shared libraries needed by an executable at run-time, just before it is executed. This is in contrast to static linking, where all code is bundled into a single executable file at compile-time. ELF files distinguish themselves as needing dynamic-linking by the presence of a special section in the file, \verb`.interp` (interpreter), that holds the path of the dynamic linker to be invoked by the operating system. In Linux, when an ELF binary is executed, if the interpreter value is set, the interpreter itself is loaded first with arguments pointing to the original binary such that the interperter can perform relocations and any other dynamic-linking tasks.

Dynamic loaders are often tightly coupled to the C standard library (\verb`libc`) implementation. The most popular open source libc implementation is the GNU C library~(glibc) which is the default on most Linux distributions. Another popular standard library is musl, the default on Alpine Linux, which prides itself on being lightweight, fast, and simple. We replaced several facets of the musl dynamic loader with declarative SQL and were surprised with the clarity and conciseness that emerged.

\begin{figure*}[ht]
\centering
\begin{minipage}[b]{.49\textwidth}
   \centering
   \begin{minted}[frame=single,fontsize=\footnotesize]{c}
static size_t count_syms(struct dso *p){
  if (p->hashtab) return p->hashtab[1];
  size_t nsym, i;
  uint32_t *buckets = p->ghashtab + 4 +
        (p->ghashtab[2]*sizeof(size_t)/4);
  uint32_t *hashval;
  for (i = nsym = 0; i < p->ghashtab[0]; i++) {
    if (buckets[i] > nsym)
      nsym = buckets[i];
  }
  if (nsym) {
    hashval = buckets + p->ghashtab[0]
        + (nsym - p->ghashtab[1]);
    do nsym++;
    while (!(*hashval++ & 1));
  }
  return nsym;
}
\end{minted}
\end{minipage}\hfill
\begin{minipage}[b]{.49\textwidth}
   \centering
   \begin{minted}[frame=single,fontsize=\footnotesize]{c}
static size_t count_syms(struct dso *p) {
  char sql[1024];
  snprintf(sql, sizeof(sql), 
        "SELECT COUNT(*) FROM ELF_SYMBOLS"
        "WHERE imported = 1");
  sqlite3_stmt *stmt = execute_sqlite3_stmt(sql);
  size_t count = sqlite3_column_int(stmt, 0);
  sqlite3_finalize(stmt);
  return count;
}
\end{minted}
\end{minipage}
\caption{Rewriting the \texttt{count\_syms} function in the musl dynamic loader to make use of the SQLite database}
\label{fig:count-syms-code}
\end{figure*}

Consider the simple case of counting, iterating and finding a symbol in an ELF file. The symbols are stored in a section \emph{contiguously} such that repetitive linear searches would not be ideal. Creating a index-table (hash map) at runtime may introduce unnecessary allocations and additional work which may not be desired. ELF takes the approach of embedding additional data structures within the file format itself, purely by convention, such as a pre-computed hash map (\verb`DT_HASH`) and a bloom filter (\verb`DT_GNU_HASH`) to improve these operations. The need to know the underlying storage medium and additional data structures complicates the code and creates a leaky abstraction. Conversely, the code depicted in Figure~\ref{fig:count-syms-code} clearly articulates the desired result and goes so far as to even describe that it should be symbols that are \emph{imported} (i.e., need to be provided by an external library and present in the dynamic symbol table).

Our SQL-based implementation allowed us to express the intent of the original code better than it did, exposing a known bug~\cite{cvscommi59:online} in the GCC (compiler) toolchain that musl failed to account for. The original musl code is merely counting \emph{all} symbols present in the dynamic symbol table (\verb`.dynsym`) which incorrectly included two symbols that GCC accidentally included during object creation. Using SQL allows for a lot more expressiveness, conveying its intent to the reader \emph{and} the author.

\subsubsection{symbol interposition}

Dynamic linking is the predominant way in which software is deployed and installed on most Linux distributions. Dynamic-linking offers the advantage of reduced file system footprint since library use is consolidated largely to a single file and version. Furthermore it facilitates security updates since only that single file (library) needs to be updated. Symbol interposition is a feature of the dynamic linking process with ELF that allows the dynamic linker to resolve symbols, such as function or variable names, at runtime. A symbol name may be exported by more than one ELF shared object file, and the binding to the exact symbol depends on the dynamic linker; typically, it's dictated by the loading order. Effectively, the key-space for symbol names are flat and there is no namespacing of symbol names within a particular ELF file. The ability for symbols to interpose is what facilitates the \verb`LD_PRELOAD` functionality which allows one to easily override the resolution of a group of symbols. This capability is somewhat controversial as in order to support interposition, the compilers must emit assembly such that every function jumps to the procedure linkage table~(PLT) that results in a constant performance penalty even when the capability is unused.

\begin{figure}[t]
    \begin{minted}[frame=single,fontsize=\small]{sql}
SELECT name, version,
       count(*) as symbol_count,
       GROUP_CONCAT(path, ':') as libraries
FROM elf_symbols
WHERE exported = TRUE AND section != '.bss'
GROUP BY name, version
HAVING count(*) >= 2
\end{minted}
\caption{SQL query to audit any symbol interposition between libraries.}
\label{fig:symbol-interposition-sql}
\end{figure}

Excluding interposition that is explicit through the use \verb`LD_PRELOAD`, the presence of the same symbol name across two dynamic objects creates a risk of failure if the linking order were to be changed or a symbol removed. \sys allows one to easily identify whether such a situation has occurred and whether the interposition is a unintended but benign side-effect of the link order. Linux distributions often go through the exercise of combing through any possible symbol interposition by auditing every binary as well as the much larger cross-product of all possible library combinations. These distributions often employ one-off custom written scripts to audit potential conflicts~\cite{FC6libra44:online}.

\sys can be given multiple binaries or a directory allowing a single query to find all possible symbol interpositions using a SQL query like the one in Figure~\ref{fig:symbol-interposition-sql}. To facilitate understanding possible conflicts within a single executable, \sys can be given the \verb`--recursive` flag which will load all dependent shared objects as determined by the dynamic loader into the database.

\subsubsection{aggregate analysis}

Tools such as \verb|objdump| and \verb|readelf| aim to provide information about a single ELF file. These tools allow you to analyze various attributes (e.g. header, symbols and sections information) of a binary file. Although they are powerful for analyzing individual binaries, they do not inherently provide built-in support for aggregate analysis of multiple ELF files. 
Although it is possible to use a scripting language like Python to perform aggregate analysis on multiple ELF files, it introduces an additional layer of complexity. \verb|elf_diff|~\cite{noseglasses23:online} is a project written in Python that uses various tools such as \verb|objdump| and \verb|nm| for aggregate analysis of ELF files. However, there are some shortcomings of such a project:
\begin{itemize}
    \item \textbf{Manual Parsing Complexity}: The project involves manually parsing the output of tools \verb|objdump| and \verb|nm| in Python, leading to increased complexity in handling diverse output formats. Different versions of these tools or variations in platform-specific implementations may require careful parsing logic.
    \item \textbf{Limited Aggregation Capabilities}: The project may face limitations in terms of aggregating and correlating information across multiple ELF files. Without a centralized data structure or database, combining and comparing results from different files could be cumbersome and error-prone.
\end{itemize}

\begin{figure*}[ht]
\centering
\begin{minipage}[b]{.49\textwidth}
   \centering
    \begin{minted}[frame=single,fontsize=\footnotesize]{python}
def extractSymbols(self, filename: str) -> None:
    """Gather the properties of a symbol"""
    nm_output_mangled: str = self._readNMOutput(
        filename=filename, extra_flags=[]
    )
    nm_output_demangled: str = self._readNMOutput(
        filename=filename, extra_flags=["-C"]
    )
    nm_regex_mangled = re.compile(
        r"^[0-9A-Fa-f]+\s([0-9A-Fa-f]+)"
        r"\s(\w)\s([^\t]+)(\t(.+))?"
    )
    nm_regex_demangled = re.compile(
        r"^[0-9A-Fa-f]+\s([0-9A-Fa-f]+)\s(\w)\s(.+)"
    )
    lines_mangled = nm_output_mangled.splitlines()
    lines_demangled = nm_output_demangled.splitlines()
    for line_mangled, line_demangled in 
        # ... 9 other lines ...
        symbol_name_is_demangled: bool
        (
            symbol_name,
            symbol_name_is_demangled,
        ) = self._demangle(
          symbol_name_with_mangling_state_unknown
        )
    \end{minted}
    \end{minipage}\hfill
    \begin{minipage}[b]{.49\textwidth}
    \centering
    \begin{minted}[frame=single,fontsize=\footnotesize]{python}
def sqlExtractSymbols(self, filename: str) -> None:
  """Gather the properties of a symbol"""
  result = list(
    self.engine.execute(
      """SELECT name, demangle_name, type, size
         FROM ELF_SYMBOLS
         WHERE path = :path AND size != 0 AND
         (type= 'FUNC' OR type = 'OBJECT')""",
      {"path": filename}
    ))
    \end{minted}
    \end{minipage}
    \caption{rewrite \texttt{extractSymbols} function in \texttt{elf\_diff}}
\label{fig:symbol-extraction-elf-diff}
\end{figure*}
 These limitations can be overcome by using \sys. The function utilized by \verb|elf_diff| for symbol extraction is illustrated on the left in figure~\ref{fig:symbol-extraction-elf-diff}. The code initially invokes the \verb|_readNMOutput| helper method to execute the \verb|nm| command with various flags. Then, it extracts demangled and mangled symbols. Subsequently, regular expressions are employed to parse the output from \verb|nm| to handle both kinds of symbol names. The code then executes a series of conditional statements to determine symbol properties. The whole process appears to be counterintuitive and intricate. Conversely, utilizing \sys would streamline this process with a single, straightforward SQL query, which is illustrated on the right of figure~\ref{fig:symbol-extraction-elf-diff}. The original code involves manual parsing of \verb|nm| output using regular expressions, which can be error-prone and complex. Utilizing an SQL query to directly extract pertinent information from a relational data model simplifies the code. Furthermore, we have reduced the code length from 59 lines to 28 lines, improving code readability and maintainability in contrast to intricate parsing logic.

In addition, for aggregation capabilities, \verb|elf_diff| use various data structures to store binary, symbol, and instruction information separately. While this approach may be suitable for small binary files, it exhibits limitations when it comes to contextual analysis. The lack of a centralized structure makes it challenging to identify patterns, trends, or anomalies that might emerge only when considering the collective data from multiple ELF files. By contrast, using the data model implemented by \sys can significantly enhance the project's ability to perform nuanced contextual analysis. With data stored in a centralized database, patterns and trends that span across multiple ELF files become more easily identifiable and analyzable. This centralized storage solution facilitates a more holistic understanding of the data, enabling the project to uncover insights that might be challenging to discern when relying on disparate data structures.

\begin{figure}
    \centering
    \includegraphics[width=\columnwidth]{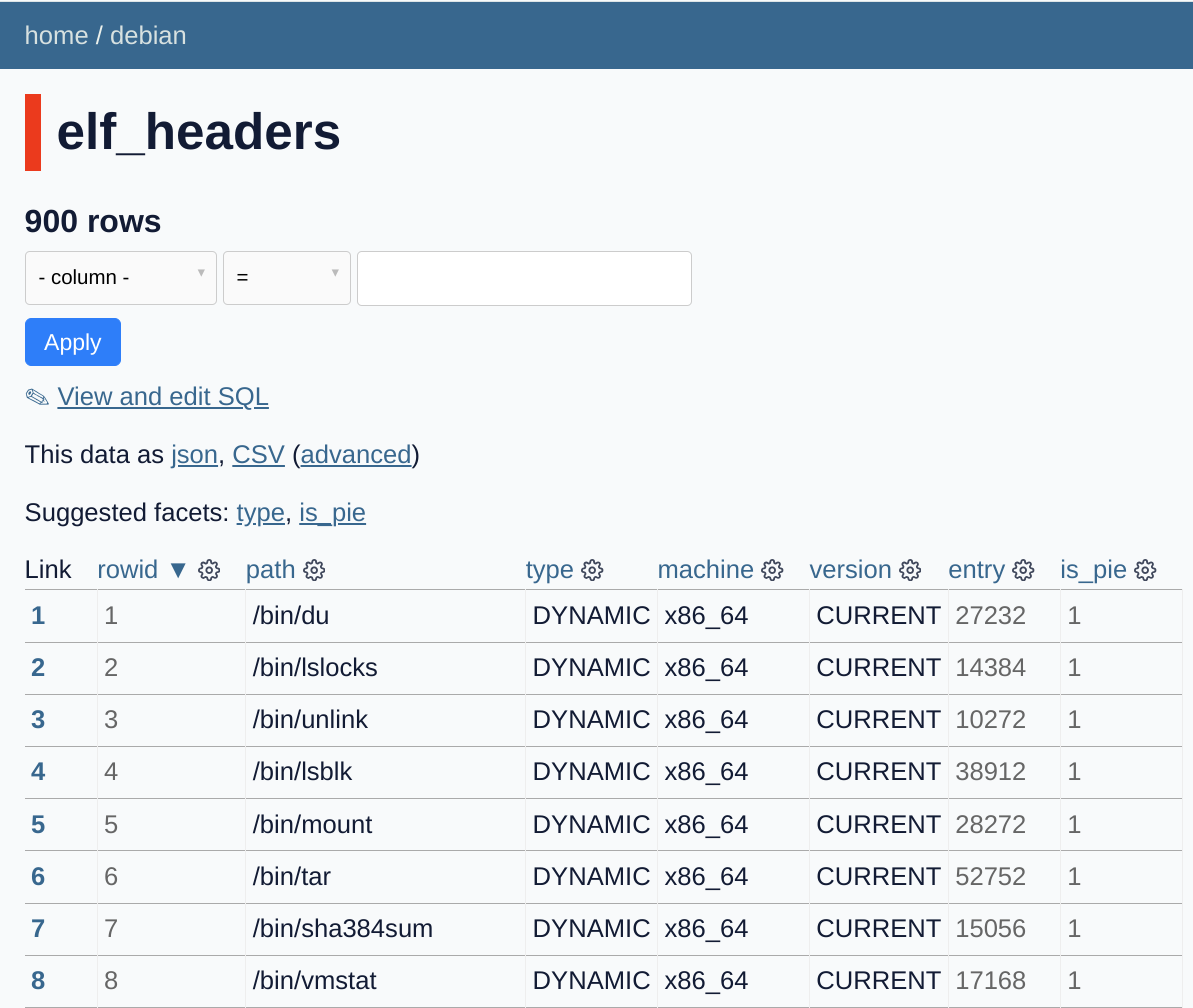}
    \caption{Leveraging the Datasette tool to explore a snapshot of a complete Debian release. }
    \label{fig:datasette-debian}
\end{figure}

Aggregate analysis need not be restricted to binaries on the order of several, but can constitute a complete distribution as well. A common task amongst distribution maintainers is to audit the list of symbols amongst all possible installable shared libraries and check for conflicts. Any duplicates that are found are then investigated further to determine if they could belong to the same executable's dependency graph; such a possibility would likely indicate a bug and possible failure. This task is often relegated to \emph{one-off} scripts and back-and-forth discussions on mailing lists~\cite{FC6libra44:online}. Using \sys we were able to take a complete snapshot of all executables and shared objects on a fresh Debian (12.0) installation memoized into a single SQLite database. We have made this database available online for interactive exploration via Datasette, a popular project in the SQL ecosystem as depicted in Figure~\ref{fig:datasette-debian}.

\subsection{Performance Evaluation}
\label{sec:evaluation:performance}\label{sec:eval:perf}

\begin{figure}[htb!]
    \centering
    \includegraphics[width=\columnwidth]{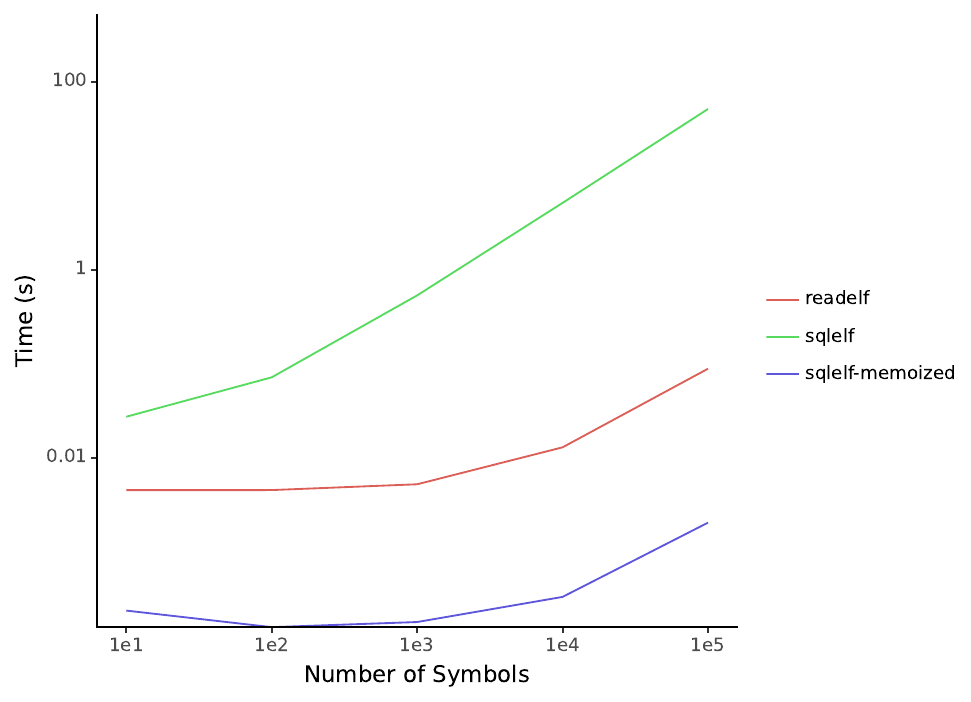}
    \caption{\sys vs. readelf in discovering number of symbols}
    \label{fig:benchmark-symbol-size}
\end{figure}

Python was chosen as the development language for its ease of use, iterative speed, and widespread audience. Unsurprisingly however, the choice of language comes at a performance cost when compared to the traditional tools (\verb`readelf` or \verb`objdump`) which are written in C/C++. Since any table may be a bottleneck for \sys, we can use the single table \verb`ELF_SYMBOLS` as a benchmark. Additionally, by offering a SQL front-end, the performance is also affected by the complexity of the task and query. To establish some ground-truths, we chose to look at a straightforward query that can be easily replicated using \verb`readelf` and \verb`wc`: count the number of symbols in an ELF file. We benchmarked the contrasting pipelines by generating executable ELF files with the desired number of symbols. Although consistently slower than its counterparts, the tool performs acceptably well for ELF files with up to 10,000 symbols achieving sub-second execution times. As the number of symbols begins to approach and go beyond $1\mathrm{e}{5}$ the latency drastically increases as depicted in Figure~\ref{fig:benchmark-symbol-size}. To address this, we have added support in the tool to export the in-memory SQLite database to a file on disk. Memoizing the creation of the database in a way that allows other SQLite tools to access it allowed us to use the native C/C++ SQLite client and avoid our Python implementation's overhead. As shown in Figure~\ref{fig:benchmark-symbol-size}, this tool consistently outperformed the \verb`readelf` pipeline.

\begin{figure}[htb]
    \centering
    \includegraphics[width=\columnwidth]{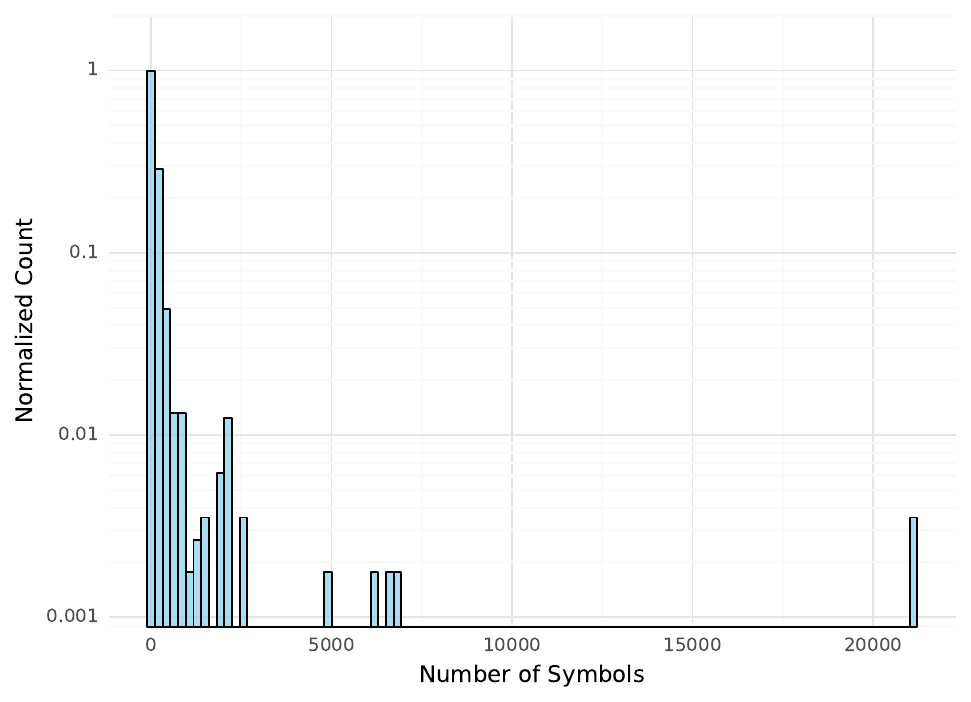}
    \caption{Number of symbols per executable on a typical Ubuntu installation}
    \label{fig:ubuntu-symbol-histogram}
\end{figure}

We were interested in understanding how realistic are binaries that exceed $1\mathrm{e}{4}$ symbols. We audited roughly 1600 ELF files (\verb`/bin`, \verb`/usr/bin` and \verb`/lib/x86_64-linux-gnu`) on a fresh Ubuntu 22.04 installation and failed to find a binary whose symbol count surpassed 30,000 shown in Figure~\ref{fig:ubuntu-symbol-histogram}. The largest ELF file in terms of symbol was the tool \verb`lto-dump` with 21127 symbols. Nearly all installed binaries on a Debian based machine are dynamically-linked which would reduce the maximum number of symbols per any given file. At Google, most binaries are statically-linked and it is not uncommon to find extremely large binaries (>2GiB). It's common for such binaries to have well over 1 million ($1\mathrm{e}{6}$) symbols. With so many symbols, even generating the memoized database can be an overly time-consuming step because \sys still must rely on Python for this. 

%% file: text/related.tex
\section{Related Work}\label{sec:related}

\sys is the first tool to support a declarative model for analyzing the object metadata throughout an entire system. 

Recent systems have explored the challenges in managing software dependencies with shared libraries.  Most of this work focus on isolating bugs in dependent libraries to reduce \emph{supply chain vulnerabilities}\cite{Cadariu15}.  For example, \textsc{BinWrap}~\cite{Christou23} isolates javascript dependencies by creating new permission models and \textsc{Harp}~\cite{Vasilakis21} uses a learning process to eliminate malicious behavior from a dependency.  These tools aid with attack prevention, but do not help a system administrator with the fundamental task of managing the objects in their system.  iFed~\cite{Ren22} proposes a new dynamic loader design to accelerate the process of loading binaries; it does not aid with the challenges in managing objects in a system.

The rest of this section describes similar efforts to use declarative models for reasoning about computer systems. 

Numerous systems adopt declarative languages, including relational languages, to simplify dynamic analyses, including Fay~\cite{Erlingsson11}, EndoScope~\cite{Cheung08}, PQL~\cite{Martin05}, PTQL~\cite{Goldsmith05}, PivotTracing~\cite{Mace15}, \textit{asmdb}~\cite{ayers2019asmdb}  and the OmniTable~\cite{Quinn22}.  Such systems make it easier for developers to track the execution behavior of their applications, which is important for system administration.  However, they do not simply the task of managing shared libraries themselves and are thus complimentary to \sys. 

Other systems adopt declarative languages to support static analysis, including PQL~\cite{Martin05}, bddbdbd~\cite{Whaley04}, Flix~\cite{Madsen16}, and Datafun~\cite{Arntzenius16}.  Such tools focus only on the code portions of a shared object and ignore the complexity inherent in managing shared objects in a large system.  \sys is complementary to such tools. 

Other system administration tools have incorporated declarative languages and SQL to empower users.  For example, Nushell~\cite{Nushell33:online} offers a structured approach to data manipulation and visualization for interacting with traditional linux binaries.  \sys and nushell share a similar vision---to empower system administration by structured abstractions---and compose together nicely.

%% file: text/conclusion.tex
\section{Conclusion}\label{sec:conclusion}
We presented \sys and associated data model as a solution to help developers analyze the (meta)data of their system.
We show the usefulness of the model on \numcases case studies that illustrate: ease of use, generality and simplification over existing methadologies.

\sys is available under the permissive MIT open-source license and can be found on GitHub~\footnote{We will de-anonymize post-review}. Additionally, \sys is packaged and available on Python's standard packaging index PyPI. It can be easily installed via standard Python tooling (\eg \texttt{pip}). We welcome issues, bug-fixes and contributions.